# PECULIARITIES OF THE SERS SPECTRA OF 4,4'-BIPYRIDINE MOLECULE IN A SINGLE MOLECULE DETECTION REGIME

## A.M. Polubotko[1], E.V. Solovyeva[2]

[1]A.F. Ioffe Physico-Technical Institute, Politechnicheskaya 26, 194021 Saint Petersburg Russia, E-mail: alex.marina@mail.ioffe.ru

[2]Saint Petersburg State University, The Institute of Chemistry, The University pr. 26, 198504 Petrodvoretz, Saint Petersburg, Russia,
E-mail: solovyeva.elena.v@gmail.com

## Abstract

The paper briefly describes main statements of the theory of the SERS spectra with regards to the single molecule regime, when the enhancement achieves the values $\sim 10^{14} - 10^{15}$. Analysis of the spectra of 4,4'-bypyridine, obtained on the dimer lattice of sharp nanoparticles points out that the observed enhancement is caused exclusively by a strong quadrupole light-molecule interaction, which manifests in the presence of lines, caused by vibrations with the unit irreducible representations of the $D_2$ and $D_{2h}$ symmetry groups, which apparently describe the symmetry properties of the molecule. The study of the spectra, obtained by Tip enhanced spectroscopy demonstrates that the strong quadrupole light-molecule interaction still plays a leading role, however the strong dipole interaction still manifests in the existence of very weak forbidden lines. This result apparently is associated with another experimental geometry
.

## 1. Introduction

Investigation of optical phenomena on molecules, adsorbed on metal surfaces is of great interest both from fundamental and applied points of view. A study of the SERS spectra in the regime, when one can be able to detect a small amount, or even a single molecule [1,2] is of particular interest. Two mechanisms of SERS are accepted at present:: the electrodynamical one, which causes the value of the enhancement $\sim 10^4$, and a "chemical mechanism", which manifests only in the first layer of adsorbed molecules with the additional enhancement $\sim 10^2$. The electrodynamical mechanism one usually associates with the enhancement of so-called surface plasmons. In [3,4] it was demonstrated that the idea of the "chemical mechanism" is mistaken, the additional enhancement $\sim 10^2$ is of a pure electrodynamical nature and associated with a very large difference in the enhancement of the electric field and its derivatives in the first and the second layers of adsorbed molecules in the regions of the metal surface with a very large positive curvature. It is so-called rod effect. At present it is a base of a new spectroscopic method - Tip enhanced Raman spectroscopy (TERS). In addition we demonstrated ([3] for example) that the huge enhancement in SERS is associated not only with the enhancement of the electric field, but with existence of so-called strong quadrupole light-molecule interaction, associated with a huge increase of the electric field derivatives $\frac{\partial E_\alpha}{\partial x_\alpha}$, and with an unique role of the quadrupole moments with the same indices $Q_{exx}, Q_{eyy}$ and $Q_{ezz}$. The quadrupole interaction can be stronger, than the dipole one and can play a leading role in the enhancement processes. In [5], it was demonstrated with the help of a crude estimation that the SERS enhancement in the single molecule regime is caused exclusively by the strong quadrupole light-molecule interaction, while the enhancement of the electric field and the enhancement of the dipole interaction have not so valuable role. In the present work we shall demonstrate that the result obtained in [5] is confirmed by the analysis of the SERS spectra of such symmetrical molecule as 4,4'-bipyridine obtained in proper conditions. However, in the case of the TERS technic application for the single molecule detection, the dipole interaction can be sufficiently strong with respect to the quadrupole one and nonetheless it manifests in the spectra.

## 2. Main statements of the dipole-quadrupole SERS theory

The Dipole-Quadrupole theory of surface enhanced optical processes is published in detail in our works. The SERS theory is published in the monograph [3], the SEHRS theory is presented in [6] and the SEIRA theory in [7]. Therefore, here we expound only main statements, which are important for this work.

As it is well known, the light-molecule (light-electron) interaction Hamiltonian has the following general form.

$$\widehat{\mathbf{H}}_{e-r} = -\sum_i \frac{ie\hbar}{mc} \mathbf{A}_i \nabla_i \tag{1}$$

Here summation is over all electrons, $\mathbf{A}_i$ is a vector potential of the electromagnetic field at the place of the $i$ electron. Other designations are conventional. Taking into account small dimensions of the molecule and expanding the vector potential into the power series, the Hamiltonian (1) can be transferred to the form

$$\widehat{\mathbf{H}}_{e-r} = |\mathbf{E}| \frac{(\mathbf{e}^* \mathbf{f}_e^*)e^{i\omega t} + (\mathbf{e}\mathbf{f}_e)e^{-i\omega t}}{2}, \tag{2}$$

where $\mathbf{E}$ is the electric field, $\mathbf{e}$ is a polarization vector, $\omega$ is the frequency of the electromagnetic field, $\mathbf{f}_e$ is a generalized vector of light-electron interaction with the components

$$f_{e\alpha} = d_{e\alpha} + \frac{1}{2E_\alpha} \sum_\beta \frac{\partial E_\alpha}{\partial x_\beta} Q_{e\alpha\beta} . \tag{3}$$

Here $d_{e\alpha}$ and $Q_{e\alpha\beta}$ are respectively $\alpha$ and $\alpha\beta$ components of the dipole moments and of the quadrupole moments tensor.

As it is well known from usual spectroscopy, one usually takes into account only the dipole interaction since the quadrupole terms are very small. However, since the interaction in surface enhanced spectroscopy arises in a surface field but not in the field in a free space, the characteristic length of the change of the electric field is $l$, which has a sense of the characteristic size of the roughness. The role of the quadrupole interaction strongly increases because of a strong increase of the derivatives $\frac{\partial E_\alpha}{\partial x_\alpha}$ and due to a unique role of the quadrupole moments $Q_{e\alpha\alpha}$ with a constant sign. This situation occurs since the matrix elements of these moments, which define such physical values as the cross-sections of scattering or absorption must be significantly larger than the matrix elements of the moments $Q_{e\alpha\beta}$ ($\alpha \neq \beta$), since the integrands are "more oscillating". This effect is manifestation of some "quantum interference" because it is associated with the necessity of the quantum mechanical averaging and calculation of matrix elements of the quadrupole moments operators with some spatial oscillation, which depends on the sign and becomes larger for operators with the changeable sign that must result in a strong decrease of their values. Further we shall name the moments $Q_{e\alpha\alpha}$ as main quadrupole moments $Q_{main}$. The same one can say concerning the dipole interaction. Since the dipole moments are the values of a changeable sign, then the relative role of the quadrupole interaction with the moments $Q_{e\alpha\alpha}$ strongly increases with respect to the dipole interaction with the moments $d_\alpha$ in comparison with considered earlier.

As it is well known ([2] for example), there is a strong enhancement of the electric field component $E_z$, which is perpendicular to the surface and its derivative $\frac{\partial E_z}{\partial z}$ near such features of the metal rough surface as the areas with a large positive curvature. For example, near such a roughness model as a tip, the radial component of the electric field $E_r$ has the form

$$E_r \sim |\mathbf{E}_{inc}| C_0 \left(\frac{l_1}{r}\right)^{\beta}, \tag{4}$$

and $E_r \to \infty$ when $r \to 0$ ($r$ - is the distance from the top of the tip). Here in (4), $|\mathbf{E}_{inc}|$ - is the amplitude of the incident electric field, $C_0 \sim 1$ is some numerical coefficient, associated with the geometry of the experiment, $l_1$ - is a characteristic size of the tip, $0 < \beta < 1$ and depends on the tip angle and the complex dielectric constant of the tip.

For estimation of the relative role of the quadrupole interaction with respect to the dipole one, one can take the value

$$\frac{\overline{\langle n|Q_{e\alpha\alpha}|l\rangle}}{\overline{\langle n|d_{e\alpha}|l\rangle}} \frac{1}{2E_{inc}} \frac{\partial E_\alpha}{\partial x_\alpha} \approx Ba \frac{1}{E_{inc}} \frac{\partial E_\alpha}{\partial x_\alpha} \tag{5}$$

where

$$Ba = \frac{\overline{\langle n|Q_{e\alpha\alpha}|l\rangle}}{\overline{\langle n|d_{e\alpha}|l\rangle}} \tag{6}$$

is a relation of some mean values of matrix elements of the quadrupole and dipole moments, $a$ is the size of the molecule, $B \gg 1$ is some numerical coefficient, associated with the fact that the quadrupole moments $Q_{e\alpha\alpha}$ are of constant sign and the dipole moments $d_{e\alpha}$ are of a changeable sign.

A crude estimation for such molecules as benzene, pyridine and pyrazine gives the value $B \sim 2 \times 10^2$ [2] and for the crystal violet molecule $B \sim (1.15 \times 10^2 - 2.7 \times 10^2)$ [5]. One should note that in accordance with (6) it is sufficiently crude estimation, which depends on $\alpha$. However, the fact that it is a large value $\sim 10^2$ is sufficient. Taking into account that SERS is the process of the second order and depends from two processes, absorption and spontaneous emission of the light, the enhancement in SERS due to a pure dipole interaction is

$$G_d \sim C_0^4 \left(\frac{l_1}{r}\right)^{4\beta}, \qquad (7)$$

and due to the pure quadrupole interaction is

$$G_Q \sim \beta^4 C_0^4 \left(\frac{B}{2}\right)^4 \left(\frac{l_1}{r}\right)^{4\beta} \left(\frac{a}{r}\right)^4 . \qquad (8)$$

Their relation is

$$\frac{G_Q}{G_d} \sim \beta^4 \left(\frac{Ba}{2r}\right)^4 . \qquad (9)$$

One can see from (7, 8) that both the pure dipole and the pure quadrupole enhancement are proportional to the $4\beta$ power of the tip dimension $l_1$ and to the forth power of the size of the molecule $a$. The more these values, the more the enhancement. One should note that this is a consequence of a quasistatic approximation and is in a good agreement with the experimental observation of single molecules, which were observed on sufficiently large colloidal particles with a characteristic size $\sim 100 \, nm$ and on sufficiently large molecules such as R6G and crystal violet. The enhancement of the dipole and especially of the quadrupole interaction strongly increases, when one moves towards to the top of the tip. For sufficiently small $r$

$$r < \beta \frac{Ba}{2}, \qquad (10)$$

the quadrupole interaction becomes larger, than the dipole one. From (7-10) one can demonstrate that for the tip with the dimensions $l_1 \sim 100 nm$, $B \sim 2 \times 10^2$ $a \sim 0.5 nm$, $C_0 \sim 1$, $\beta \approx 1$ the enhancement due to the quadrupole interaction $\sim 10^{14}$ can be achieved for $r \sim 1.26\ nm$, that is a reasonable value, when both the values of the field and its derivatives apparently are reasonable and can be achieved in an experiment. The enhancement due to the pure dipole interaction is about $2.4 \times 10^6$ times less, than one due to the quadrupole interaction. One can see that taking into account a strong distance dependence of the quadrupole interaction $\sim r^{-4-4\beta}$, one can compensate the reasonable change of the parameters $C_0$ and $\beta$ by a small change of the $r$ value. The $r$ value remains in a reasonable interval in order to obtain the enhancement value due to the quadrupole interaction $\sim 10^{14}$ .and the enhancement, caused by the dipole interaction can be neglected.

As it was demonstrated in [2], the cross-section for symmetrical molecules in SERS is determined by the sum of various scattering contributions, which occur via various combinations of the dipole and quadrupole moments $T_{(s,p), f_1 - f_2}$ and which has the form

$$d\sigma_{SERS_s} \sim \sum_p \sum_{f_1, f_2} T_{(s,p), f_1 - f_2} . \qquad (11)$$

In [2] one can find an evident expression for the cross-section, which is very cumbersome and therefore is omitted here.

In (11), $(s, p)$ are quantum numbers of corresponding degenerate mode, $s$ is a quantum number, which numerates the group of degenerate states, $p$ numerates the states inside the group. The contributions $T_{(s,p), f_1 - f_2}$ determine individual scattering processes, which occur via various combination of the dipole and quadrupole moments $f_1$ and $f_2$ (Figure 1). In accordance with [2], these contributions obey selection rules

$$\Gamma_{(s,p)} \in \Gamma_{f_1} \times \Gamma_{f_2}, \qquad (12)$$

where $\Gamma$ designates the irreducible representation, which describes transformational properties of the vibrational mode $(s, p)$ and the dipole and quadrupole moments $f_1$ and $f_2$. Here we consider the 4,4'- Bipyridine molecule, which can be described by the symmetry groups $D_2$ or $D_{2h}$, with one-dimensional irreducible representations and therefore degeneracy is absent. The quadrupole moments $Q_{exx}, Q_{eyy}$ and $Q_{ezz}$ transform after the unit irreducible representation both in the $D_2$ and $D_{2h}$ symmetry groups.

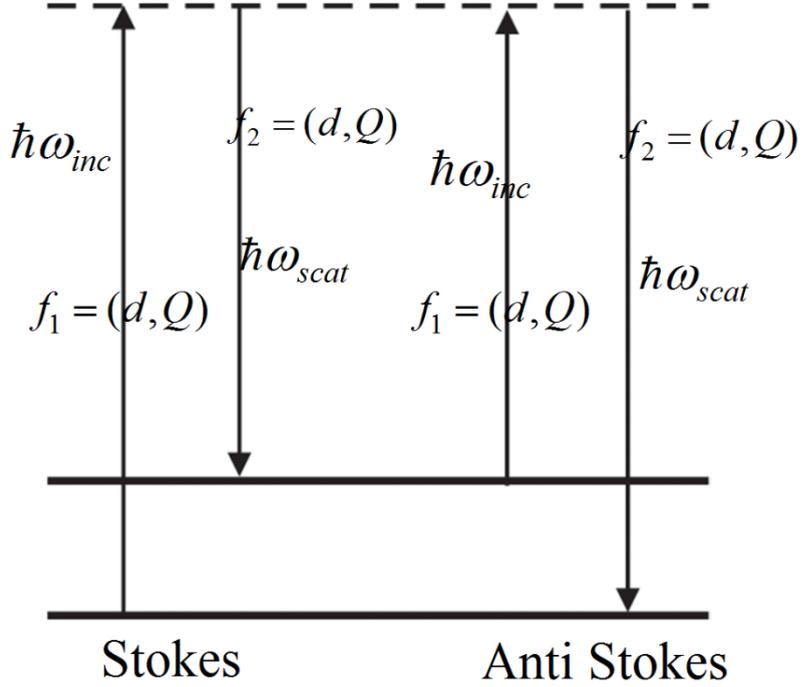

Figure 1. The scattering diagram for SERS. The scattering can occur via various combinations of the dipole and quadrupole moments $d$ and $Q$.

The case of single molecule detection refers to the one of a very strong enhancement, when the quadrupole interaction with the main quadrupole moments must be significantly stronger, than the enhanced dipole interaction. Let us designate the contributions into the cross-section $T_{(s,p), f_1-f_2}$ simply as $(f_1 - f_2)$. In accordance with the above ideas the most enhanced contributions are those, which are caused by the scattering on the two main quadrupole moments, or of the $(Q_{main} - Q_{main})$ type. The contributions $(Q_{main} - d_\alpha)$ caused by the scattering via main quadrupole and main dipole moments and the contributions of $(d_\alpha - d_\beta)$ and $(d_\alpha - d_\alpha)$ types in particular must be enhanced also, but in a less degree. Here, in the case of single molecule detection we must consider all the dipole moments, since the molecules in these experiments are in a free state in a solution with a very small concentration [1,2], but not in an adsorbed state. The orientation in this case can be arbitrary and the enhanced electric field has projections on all coordinate axis associated with the molecule and all the terms of the dipole interaction Hamiltonian and also the contributions of the $(Q_{main} - d_\alpha)$ and $(d_\alpha - d_\beta)$ types must be enhanced. However, since in the case of the detection of a small amount of molecules or single molecule detection, the strong dipole interaction is significantly less than the strong quadrupole one, therefore, the above contributions will be significantly less. In accordance with the selection rules (12), since $Q_{main}$ transforms after the unit irreducible representation, the most enhanced contributions define the most enhanced lines, which refer to the vibrations with the unit irreducible representation. The contributions of the $(Q_{main} - d_\alpha)$ type refer to the irreducible representations, which describe transformational properties of the components of the dipole moment $d_\alpha$. One should note that the contributions of the $(d_\alpha - d_\beta)$ type and of the $(d_\alpha - d_\alpha)$ type in particular, which also must define the intensities of the lines, caused by the vibrations with the unit irreducible representation must be small and we will not take into account them further. Usually in SERS, the contributions of the ($Q_{main} - d_\alpha$) types determine appearance of forbidden lines, which are active in infrared absorption. Since these contributions are significantly less in our case, than the contributions of the ($Q_{main} - Q_{main}$) type, these lines must be practically absent. One should note that the above ideas are applicable for consideration of single molecule detection on nanoparticles, which are situated far one from another and have sharp points, when we can to model the nanoparticles by the tip. In [11] the single molecule detection was performed by the TERS technic. Therefore, apparently the final results of the spectra measurements and our conclusions differ slightly.

## 3. Analysis of the SERS spectra of 4,4'-Bipyridine for the detection of a small amount of molecules or in a single molecule regime

Overall analysis of the SERS, SEHRS, RS and IR absorption spectra of the 4,4'-Bipyridine molecule was performed in [9]. As we indicated above, there are two points of view on the structure of this molecule at present. In accordance with [12] it consists of two benzene rings, which are connected each other via the $C-C$ bond, and the opposite carbon and hydrogen atoms are substituted by nitrogen atoms. These rings are turned with respect to each other on the angle of 38.7 degrees (figure 2) and its symmetry group is $D_2$. In accordance with another opinion these rings are in the same plane and the symmetry group of the molecule is $D_{2h}$. One should note that further we shall point out the values of wavenumbers, which are observed in experimental spectra. The position of the lines in these spectra can change within the interval $\sim (10-20)$ $cm^{-1}$. Therefore, further we point out the same irreducible representation for some lines, which change their position in the indicated interval.

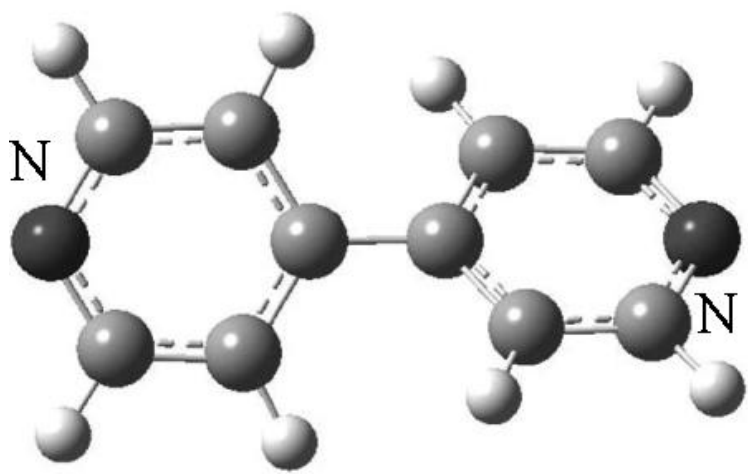

Figure 2. Possible geometry of 4,4'- bipyridine. The rings are turned on the angle 38.7 degree and the symmetry group of the molecule is $D_2$.

In accordance with the results of [8], it was demonstrated that both in the SERS and SEHRS spectra, the most part of lines can be assigned to the vibrations both with the unit irreducible representations $A_1$ or $A_g$, and with ones $B_1$ and $B_{1u}$. It is associated with peculiarities of the geometry of the molecule, when its two rings can vibrate in the same, or opposite phase. These lines have approximate values of the wavenumbers. 396, 1018, ~1064-1080, 1208, 1609 $cm^{-1}$. The frequencies of these vibrations are very close and these states are "nearly degenerate" symmetric and anti symmetric states and, therefore, they cannot be distinguished in the experiment. However, one should point out that in the case of the detection of a small amount of molecules, or single molecule detection, the contributions associated with the vibrations with the $B_1$ and $B_{1u}$ symmetry are small and therefore these lines are determined only by the vibrations with the unit irreducible representation. The lines with the wavenumbers 736, 1300 и 1516 $cm^{-1}$ refer only to the vibrations with the unit irreducible representations $A$ or $A_g$, while the lines with the wavenumbers 864, 1038 и 1487 $cm^{-1}$ refer only to the irreducible representations $B_1$ or $B_{1u}$, which describe transformational properties of the component of the dipole moment $d_z$, which is perpendicular to the surface. (Here the z axis connects the opposite nitrogen atoms). The spectrum of 4,4'-bipyridine, adsorbed from solution on the dimer lattice of silver nanoparticles (figure 3), when the authors considered that they dealt with a small amount, or even with single molecules, was investigated in [9] (figure 4). One should note that the use of the dimer roughness from our point of view is not a matter of principle. The essential point is existence of sharp points, where the field behavior is close to the singular one. In [9] one investigated a narrow interval of wavenumbers $(960-1360)$ $cm^{-1}$. Therefore, only the following lines, which refer to the vibrations transforming as the $d_\alpha$ component of the dipole moment are in this interval.

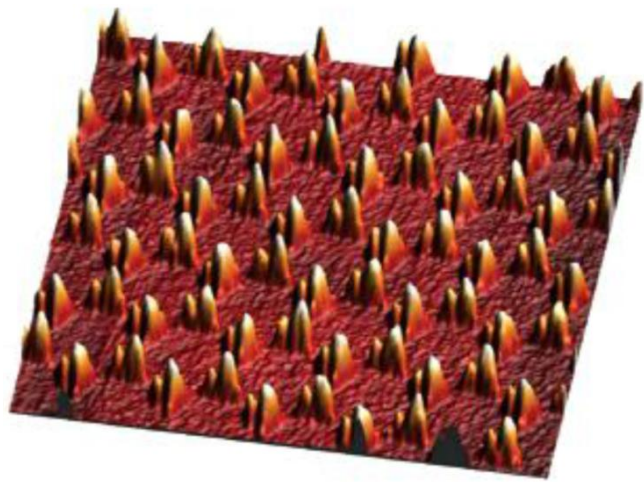

Figure 3 Typical dimer silver nano dimensional lattice [9].

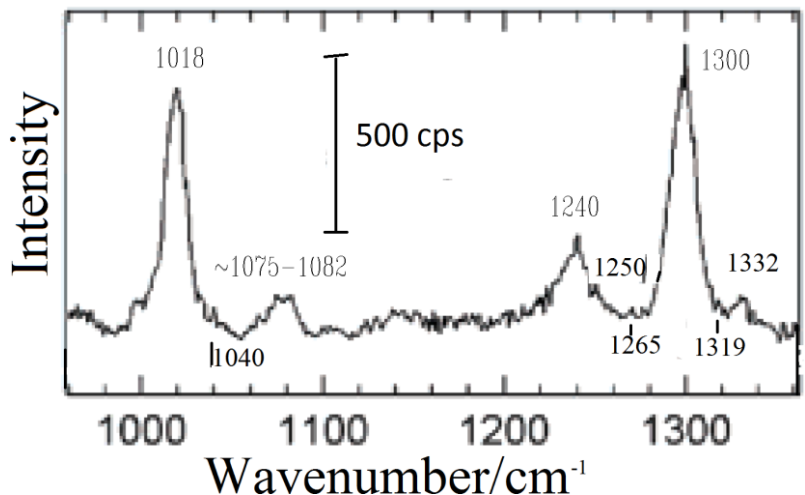

Figure 4. The SERS spectrum of a small amount of 4,4'-bipyridine molecules, adsorbed on the dimer silver lattice [9].

The lines are 1040 $см^{-1}$, which refers to the irreducible representations $B_1$ or $B_{1u}$, 1250 и 1319 $см^{-1}$, which refer to the irreducible representations $B_2$ or $B_{2u}$ and 1265 and 1332 $см^{-1}$, which refer to the irreducible representations $B_3$ or $B_{3u}$. One can see from the figure 4 that these lines, which refer to irreducible representations, which describe transformational properties of the dipole moments are very weak. This result indicates that the dipole interaction, which still manifests in this system is still sufficient in order to observe these lines, since their intensity is determined by averaging over sufficiently large volume, where the strong dipole interaction can be compared with the strong quadrupole one. However, their small intensity indicates that the strong quadrupole interaction is significantly stronger, than the dipole one in some regions of the surface. This result well corresponds to our ideas and there are strong lines with the wavenumbers 1018, ~1080 (1064), ~1218-1240, 1298-1300 $см^{-1}$, where there are the contributions, which refer to the vibrations with the unit irreducible representations $A$ or $A_g$. The same result was obtained in [10] (figure 5), where the authors observed the lines at 1018, 1078, 1240 and 1300 $cm^{-1}$ in the SERS spectra in the single molecule detection regime on the lattice with sufficiently well isolated nanoparticles. Meanwhile the lines, which refer to the irreducible representations $B_1, B_2, B_3$ or $B_{1u}, B_{2u}, B_{3u}$ respectively, are absent such as in the first case. These results confirm the fact that the enhancement in SERS in the single molecule detection regime is caused exclusively by the strong quadrupole light-molecule interaction. However, it is necessary to note that this conclusion is valid for nanoparticles with sharp points and which are in a "sufficiently large distance" one from another. In [11], the authors used the TERS technic, when the gold tip approaches to the gold plain surface on the distance, which approximately equals to the size of the molecule. The molecule in this case becomes in the field of some capacitor. Measurement of the spectra were performed in a wider range of wavenumbers from 900 till 1600 $см^{-1}$. In spite of the most enhanced lines in this case were observed at 1011, ~1078, 1218, 1291, 1516 и 1609 $cm^{-1}$, and were caused by the vibrations with the irreducible representation $A$ or $A_g$, one can see a very weak line at 1532 $cm^{-1}$, which can be assigned to the irreducible representation $B_2$ or $B_{2u}$ (figure 6). This fact indicates that in spite of the quadrupole interaction plays a major role in the enhancement in this case. However, the strong dipole interaction is sufficiently strong that causes appearance of the above line.

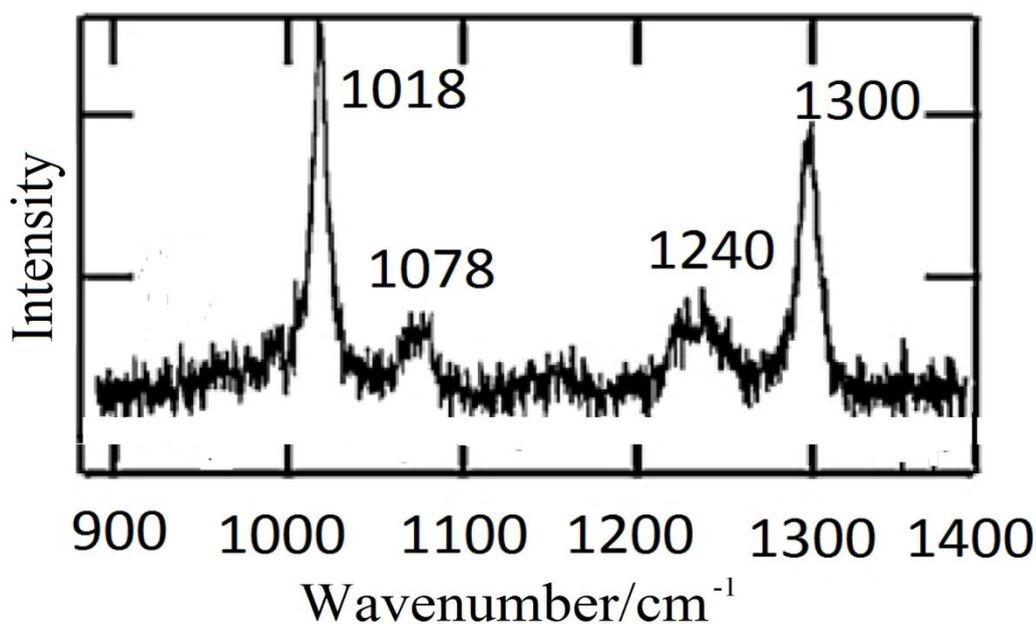

Figure 5. The Single molecule SERS spectrum of 4,4'-bipyridine, adsorbed on the dimer lattice of silver nanoparticles. [10].

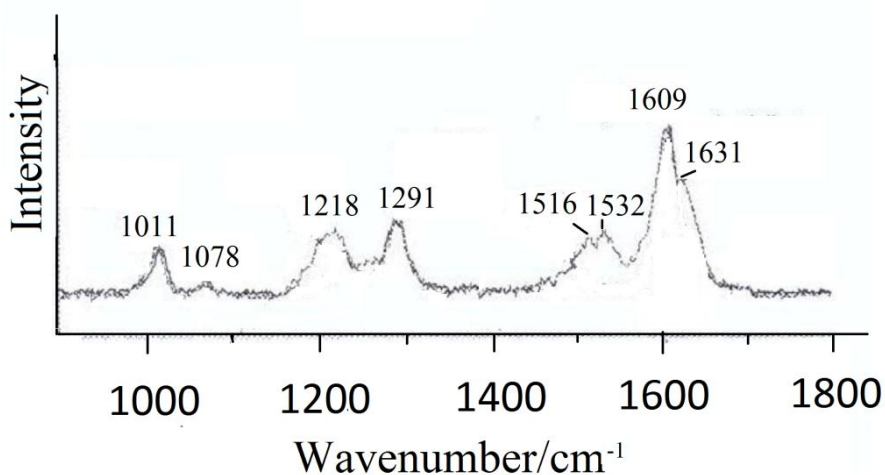

Figure 6. The spectrum of a single 4,4'-Bipyridine molecule placed between the gold tip and a gold plain surface in the TERS geometry.

## 4. Conclusion

Thus investigation of the SERS spectra of 4,4'-bipyruidine for a small amount, or single molecule detection on nanoparticles, which are sufficiently far one from another, indicates that practically all the lines, which were observed in the spectra, are caused by the contributions associated with the strong quadrupole light-molecule interaction and with totally symmetric vibrations transforming after the unit irreducible representation $A$ or $A_g$ of the corresponding $D_2$ or $D_{2h}$ symmetry group, which apparently describe the symmetry of the molecule. In addition the lines, associated with the quadrupole-dipole scattering of the $(Q_{main} - d_\alpha)$ type and with the vibrations, which transform after the irreducible representations $B_1$ or $B_{1u}$, $B_2$ or $B_{2u}$ and $B_3$ or $B_{3u}$, and describe transformational properties of other dipole moments are absent, that proves a minor role of the strong dipole interaction for these experimental conditions. In case of the use of the TERS technique, in spite of the predominant role of the quadrupole interaction, the dipole interaction can manifest itself in the spectra by appearance of forbidden lines, associated with the vibrations transforming after the irreducible representations, which describe transformational properties of the dipole moments. This result apparently is associated with an essential difference of the experimental geometry in TERS, compared with the use of the nanoparticles, which are placed sufficiently far one from another.

## References:

1. S. Nie, S.R. Emory, Science **275** Issue 5303, 1102 (1997).

2. Katrin Kneipp, Yang Wang, Harald Kneipp, Lev T. Perelman, Irving Itzkan, Ramachandra R. Dasari, and Michael S. Feld, Phys. Rev. Lett. 78 № 9 1667 (1997).

3. A.M. Polubotko, Dipole-Quadrupole Theory of Surface Enhanced Raman Scattering, Nova Science Publishers. Inc. New York 2009.

4. A.M. Polubotko, V.P. Chelibanov, Optics and spectroscopy, **119** (4), 664 (2015).

5. A.M. Polubotko, J. Opt A: Pure Appl. Opt. 1 L18-L20 (1999).

6. A.M. Polubotko, V.P. Chelibanov, Optics and spectroscopy 120, (1), 86, (2016).

7. A.M. Polubotko, Optics and spectroscopy, 114, № 5, 696 (2013).

8. A. V. Golovin, A.M. Polubotko, Chemical Physics Letters, 662, 208 (2016).

9. Yoshitaka Sawai, Baku Takimoto, Hideki Nabika, Katsuhiro Ajito, and Kei Murakoshi, J. Am. Chem. Soc. 129, 1658(2007).

10. Mai Takase, Yoshitaka Sawai, Hideki Nabika, Kei Murakoshi, Journal of Photochemistry and Photobiology A: Chemistry 221, 169 (2011).

11. Zheng Liu, Song-Yuan Ding, Zhao-Bin Chen, Xiang Wang, Jing-Hua Tian, Jason R. Anema, Xiao-Shun Zhou, De-Yin Wu, Bing-Wei Mao, Xin Xu, Bin Renand Zhong-Qun Tian, Nature communications, DOI: 10.1038/ncomms1310, (2011)

12. Zhiping Zhuang, Jianbo Cheng, Xu Wang, Bing Zhao, Xiaoxi Han, Yulie Luo, Spectrochimica Acta Part A 67, 509 (2007).